\documentclass[
    ,final            % use final for the camera ready runs
  ]
  {aipproc}

\layoutstyle{6x9}

\begin{document}

\title{The shallow phase of X--ray afterglows}

%<Replace this text with PACS numbers; choose from this list:
%                \texttt{http://www.aip..org/pacs/index.html}>}
%
% 95.30.Gv 	Radiation mechanisms; polarization
% 95.30.Jx 	Radiative transfer; scattering
% 95.85.Nv 	X-ray
% 95.85.Pw 	γ-ray
% 98.70.Rz 	γ-ray sources; γ-ray bursts
%

\author{Gabriele Ghisellini}{
  address={INAF -- Osservatorio Astronomico di Brera, Merate, Italy}
}

%\author{<author2>}{
%  address={<common address for author2 and author3>}
%}
%
%\author{<author3>}{
%  address={<common address for author2 and author3>}
%  ,altaddress={<author1 address>} % additional visiting address
%}
%

\begin{abstract}
We propose that the 
flat decay phase in the first 10$^2$--10$^4$ seconds of
the X--ray light curve of Gamma Ray Bursts can
be interpreted as prolonged activity of the central engine,
producing shells of decreasing bulk Lorentz factors $\Gamma$.
The internal dissipation of these late
shells produces a continuous and smooth emission,
usually dominant in X--rays and sometimes in the optical.
When $\Gamma$ of the late shells is larger than $1/\theta_j$, 
where $\theta_j$ is the jet opening angle, we see only a portion
of the emitting surface. 
Eventually, $\Gamma$ becomes smaller than $1/\theta_j$,
and the entire emitting surface is visible.
When $\Gamma=1/\theta_j$ there is a break in the light curve,
and the plateau ends.
During the plateau phase, we see the sum of
the ``late--prompt'' emission (due to late internal dissipation), 
and the ``real afterglow'' emission (due to external shocks).
A variety of different optical and X--ray light curves is 
possible, explaining why the X--ray and the optical light curves
often do not track each other, and why they 
often do not have simultaneous breaks.
\end{abstract}

\maketitle

%%%%%%%%%%%%%%%%%%%%%%%%%%%%%%%%%%%%%%%%%%%%
%% MAINMATTER
%%%%%%%%%%%%%%%%%%%%%%%%%%%%%%%%%%%%%%%%%%%%

\section{Introduction}

The so called ``Steep--Flat--Steep" behavior \cite{gt05, nousek05} 
% (e.g. Tagliaferri et al. 2005; Nousek et al. 2005)
of the early (up to $\sim$a day) X--ray afterglow was unpredicted
before we could observe it with {\it Swift}.
It has been interpreted in several ways
(for reviews, see e.g. \cite{zhang07})
none of which seems conclusive.
The spectral slope does not change across the temporal
break from the shallow to the normal decay phase,
ruling out a changing spectral break as a viable explanation.
An hydrodynamical or geometrical nature of the break is instead preferred.
Furthermore, the X--ray and optical lightcurves often do not 
track one another (e.g. \cite{pana06, pana07})
% Panaitescu et al. 2006; Panaitescu 2007),
suggesting a possible different origin.

To solve these difficulties Uhm \& Beloborodov \cite{uhm07}
and Genet, Daigne \& Mochkovitch \cite{genet07} suggested that
the X--ray plateau emission is not due to the
forward, but to the reverse shock running into ejecta of
relatively small (and decreasing) Lorentz factors.
This however requires an appropriate $\Gamma$--distribution of the
ejecta, and also the suppression of the X--ray flux produced by the 
forward shock.

We (\cite{gg07}) instead suggested  
that the plateau phase of the X--ray emission (and sometimes even of 
the optical) is due to a prolonged activity of the central engine (see also 
\cite{lp07}),
responsible for a ``late--prompt'' phase: 
after the early ``standard" prompt the central engine continues to 
produce for a long time (i.e. days) shells of progressively lower 
power and bulk Lorentz factor.
The dissipation process during this and the early phases 
occur at similar radii (namely close to the transparency radius).
The reason for the shallow decay phase, and for the break ending it,
is that the $\Gamma$--factors of the late shells are
motonically decreasing, allowing to see an
increasing portion of the emitting surface, until all of it is visible.
Then the break occurs when $\Gamma=1/\theta_j$.

\begin{figure}
\includegraphics[height=0.5\textheight]{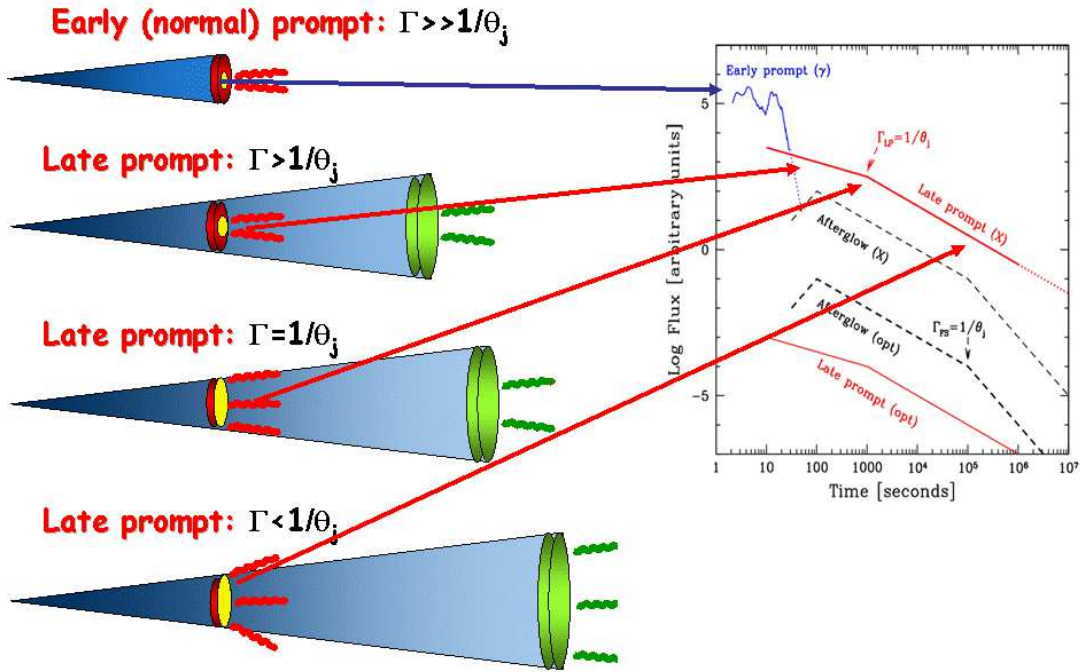}
\caption{
Cartoon of the proposed model, and schematic illustration of the different 
components contributing to the X--ray and optical light curves, as labelled.
Scales are arbitrary. 
The early prompt phase is erratic, with shells of varying $\Gamma$ and power. 
Then the central engine produces shells of progressively less power and
bulk Lorentz factors, producing a smoother light curve. 
Since the average $\Gamma$--factor is decreasing, the observer sees
an increasing portion of the emitting area, until all of it
becomes visible when $\Gamma \sim 1/\theta_j$.
When this occurs there is a break in the light curve,
associated with the ending of the shallow phase.
The case illustrated here is only one (likely the most common) 
possible case, when the X--ray flux is dominated by late 
prompt emission (solid line, the dotted line corresponds to an 
extrapolation at very late times), while the optical flux is dominated 
by the real afterglow (dashed). 
% $\Gamma_{\rm LP}$ and $\Gamma_{FS}$ indicate the $\Gamma$ of the late
% shells and the forward shocks, respectively. 
Adapted from \cite{gg07}.
}
\end{figure}

\section{The shallow X--ray afterglow phase}

\subsection{The time ending the shallow phase}

Willingale et al. \cite{willi07} have proposed to described the
X--ray afterglow light curve with a rising exponential connecting to
a power law function.
The end of the shallow phase is the junction between the exponential and
the power law, and it is called $T_a$.
They showed that interpreting $T_a$ as a jet break time one
obtains, for the {\it Swift} bursts in their sample, a good correlation
between the peak energy of the prompt spectrum, $E_{\rm peak}$,
and the collimation corrected energetics $E_\gamma$, with a small
scatter and a slope identical to the so called Ghirlanda relation \cite{ggl04}
(which identifies as a jet break time the break in the optical light curve,
occurring usually much later), challenging the physical nature of
the Ghirlanda relation.
Nava et al. \cite{nava07} have then 
investigated this issue with a larger sample, finding that the correlation
found by \cite{willi07} does not have the same slope of the
Ghirlanda one, and it is not as tight.
More importantly, they demonstrated that $T_a$ does not play any role 
in the construction of the correlation found by \cite{willi07},
which is instead (entirely) a by--product of the the $E_{\rm peak}$--$E_{\rm iso}$
correlation (the so called ``Amati" relation, \cite{ama02}).
In fact there is no (anti)--correlation between $T_a$ and $E_{\rm iso}$
(``a la Frail", \cite{frail01}) for GRBs of the same $E_{\rm peak}$
(see \cite{nava07} for more details and figures).

\subsection{Prolonged central engine activity}

The time $T_a$ is not a jet break time, still it may be produced by a 
mechanism very similar to the process responsible for the
jet break visible during the deceleration of the fireball.
Consider the accretion onto the newly formed 
black hole, and suppose that it occurs in two phases. 
The first is short, intense, erratic,
corresponding to the early prompt phase of GRBs.
The second is longer, smoother, with a rate decreasing in time,
corresponding to the late prompt emission.
The first accretion mode might correspond to the 
accretion of the equatorial core material 
which failed to form the black hole in the first place. 
It can form a very dense accreting torus, which can sustain a strong magnetic
field, which in turn efficiently extracts the rotational energy of the
black hole.  
After this phase, some fall--back material may also be accreted,
with a density  smaller than in the early phases. 
The magnetic field that this matter can sustain is weaker than
before, with a corresponding smaller power extracted from the black hole spin.
This may well correspond to production of shells of smaller 
$\Gamma$--factors.
These shells can dissipate part of their energy with the same mechanism
of the early ones. 
Occasionally, in this late prompt phase, 
the central engine may produce a faster than average shell,  
originating the late flares often observed in the Swift/XRT light curves.

In the scenario we have proposed, there is a simple relation between the 
function describing how $\Gamma$ decreases in time and the
observed decay slopes before and after $T_a$.
Assume that the plateau phase is described by $L(t)\propto t^{-\alpha_2}$,
followed by a steeper decay $L(t)\propto t^{-\alpha_3}$.
Then, by geometry alone, one can derive that (\cite{gg07}):
\begin{equation}
\Gamma \, \propto  t^{-(\alpha_3-\alpha_2)/2}
\end{equation}
We can also estimate how the barion loading of the late shells
changes in time. 
Assume $L(t) \propto \eta \Gamma\dot M c^2$, and consider
for semplicity $t>T_a$, when all the jet is visible.
Then, for constant $\eta$ we have:
\begin{equation}
\dot M \,  \propto \, t^{-(\alpha_2+\alpha_3)/2}
\end{equation}
If we insert the average values of $\alpha_3$ and $\alpha_2$ 
($\sim 1.25\pm 0.25$ and $\sim 0.6\pm 0.3$, respectively, see \cite{pana06})
we approximately have $\dot M\propto t^{-1}$ and $\Gamma\propto t^{-1/3}$.
This means that the total energy (i.e. integrated over time, 
$E =\int \Gamma \dot M c^2 dt$, beginning from the start of the plateau phase)
involved in the late phase is smaller than the energy spent during
the early prompt.

\subsection{Observational tests}

If we allow for {\it two} origins for the emission during and 
after the X--ray plateau phase (one due to the late prompt and 
the other due to the conventional forward shock), we can account 
for a variety of cases: both the optical and the X--rays
are late prompt emission or forward shock emission;
or X--rays and optical are ``decoupled'',  
one due to late prompt and the other to the forward shock.
One obvious way to check these possibilities is through the simultaneous 
spectral energy distribution (SED), which can confirm or not 
if the X--ray and the IR--optical fluxes belong to the same component.
If the emission in the two bands have a different origin
they should not ``interfere" with one another,
requiring that the X--ray spectrum breaks at low energies,
and the optical at high ($\sim$UV) energies.
The unknown extinction due to the host galaxy material
may be a complication, but infrared data can help.
The SED so obtained may clearly show if the IR--optical
and X--ray emission belong (or not) to two different components.

Since in our scenario the late central activity is not
energetically demanding, another test concerns the total kinetic energy
of the fireball after its radiative phase,
using the radio data, as done
e.g. for GRB 970508 \cite{frail00}. % , Waxman \& Kulkarni (2000).
Should the derived energetics be smaller than
what required by e.g. the refreshed shock scenario,
one could exclude this possibility, and instead
favor our scenario.

In cases in which the late prompt emission ends, the underlying 
forward shock emission can be revealed.
In the light curve, this should appear as a steep--flat transition
at late times (not to be confused with the usual
steep--flat--steep X--ray decay).
This can also be confirmed by the corresponding SEDs.

%%%%%%%%%%%%%%%%%%%%%%%%%%%%%%%%%%%%%%%%%%%%
%% Sample figure:
%%
%% The option [height=...] scales the picture to the given height,
%% without it it would be printed at its nominal size
%%%%%%%%%%%%%%%%%%%%%%%%%%%%%%%%%%%%%%%%%%%%

\begin{theacknowledgments}
I gratefully thank all my collaborators: A. Celotti, C. Firmani, G. Ghirlanda,
M. Nardini, L. Nava and F. Tavecchio.
\end{theacknowledgments}

%%%%%%%%%%%%%%%%%%%%%%%%%%%%%%%%%%%%%%%%%%%%%%%%
%% The bibliography can be prepared using the BibTeX program or
%% manually.
%%
%% The code below assumes that BibTeX is used.  If the bibliography is
%% produced without BibTeX comment out the following lines and see the
%% aipguide.pdf for further information.
%%
%% For your convenience a manually coded example is appended
%% after the \end{document}
%%%%%%%%%%%%%%%%%%%%%%%%%%%%%%%%%%%%%%%%%%%%%%%%

%%%%%%%%%%%%%%%%%%%%%%%%%%%%%%%%%%%%%%%%%%%%%%%%
%% You may have to change the BibTeX style below, depending on your
%% setup or preferences.
%%
%%
%% For The AIP proceedings layouts use either
%%%%%%%%%%%%%%%%%%%%%%%%%%%%%%%%%%%%%%%%%%%%

% \bibliographystyle{aipproc}   % if natbib is available
\bibliographystyle{aipprocl} % if natbib is missing

%%%%%%%%%%%%%%%%%%%%%%%%%%%%%%%%%%%%%%%%%%%
%% You probably want to use your own bibtex database here
%%%%%%%%%%%%%%%%%%%%%%%%%%%%%%%%%%%%%%%%%%%
% \bibliography{sample}

%%%%%%%%%%%%%%%%%%%%%%%%%%%%%%%%%%%%%%%%%%%
%% Just a reminder that you may have to run bibtex
%% All of it up to \end{document} can be removed
%% if you don't like the warning.
%%%%%%%%%%%%%%%%%%%%%%%%%%%%%%%%%%%%%%%%%%%
% \IfFileExists{\jobname.bbl}{}
%  {\typeout{}
%  \typeout{******************************************}
%  \typeout{** Please run "bibtex \jobname" to optain}
%  \typeout{** the bibliography and then re-run LaTeX}
%  \typeout{** twice to fix the references!}
%  \typeout{******************************************}
%  \typeout{}
% }

%%%%%%%%%%%%%%%%%%%%%%%%%%%%%%%%%%%%%%%%%%%
%% The following lines show an example how to produce a bibliography
%% without the help of the BibTeX program. This could be used instead
%% of the above.
%%%%%%%%%%%%%%%%%%%%%%%%%%%%%%%%%%%%%%%%%%%

% \endinput
\end{document}